\documentclass[%
 aip,
 rsi,%
 amsmath,amssymb,
reprint,%
]{revtex4-1}
\newcommand{\rchi}{Re\,$\mathit{\chi}_\perp$}
\newcommand{\ichi}{Im\,$\mathit{\chi}_\perp$}

\usepackage{graphicx}
\usepackage{dcolumn}
\usepackage{changes}
\usepackage{bm}
\usepackage{upgreek}
\usepackage[mathlines]{lineno}

\begin{document}


\title{Compact susceptometer for studies under transverse field geometries at very low temperatures}
\author{F. Rucker}
\email{felix.rucker@ph.tum.de}

\author{C. Pfleiderer}

\affiliation{ 
Physik-Department, Technical University of Munich, D-85748 Garching, Germany}

\date{\today}

\begin{abstract}
We present the design of a compact AC susceptometer for studies under arbitrarily oriented static magnetic fields, in particular magnetic fields oriented transverse to the AC excitation field. The small size of the susceptometer permits versatile use in conventional cryostats with superconducting magnet systems.  The design of the susceptometer minimizes parasitic signal contributions while providing excellent thermal anchoring suitable for measurements in a wide range down to very low  temperatures. The performance is illustrated by means of measurements of the transverse susceptibility at the magnetic field tuned quantum phase transition of the dipolar-coupled Ising ferromagnet LiHoF$_4$.

\end{abstract}

\maketitle

\section{Introduction}
Measurements of the uniform differential magnetic susceptibility $\chi$ by means of a time-dependent oscillatory excitation field represent one of the most sensitive experimental probes of the magnetic properties of condensed matter systems.\cite{1989:Deutz:RSI, 1995:Nikolo:AJP,1997_Bitko_PhD, 1997_Pfleiderer_RevSciInstrum, ppms-acs, 2004:Barbic:RSI, 2010:Yin:JLTP, 2013_Schmidt_RevScInstrum, 2015:Yonezawa:RSI, 2017:Amann:RSI, 2018_Rucker_PhD} In studies under an applied static magnetic field, the AC excitation field is conventionally oriented parallel to the applied field. However, a wide range of magnetic properties are anisotropic with respect to the underlying crystal structure and the geometry of the system. This includes anisotropic magnetic materials as well as thin films, heterostructures and nano-scaled systems. In recent years magnetic field-tuned quantum phase transitions have become topical.\cite{1996_Bitko_PhysRevLett,2011_Sachdev_Book} For these studies high sensitivity measurements of the susceptibility down to milli-Kelvin temperatures, as recorded perpendicular to the applied static field, are essential. The interest in this parameter regime implies rather stringent criteria as regards the need for a compact experimental set-up that provides excellent thermal anchoring while being suitable for measurements under large, arbitrarily oriented magnetic fields.

Different designs of AC susceptometers have been reported in the literature.\cite{1989:Deutz:RSI, 1995:Nikolo:AJP, 1997_Bitko_PhD, 1997_Pfleiderer_RevSciInstrum, ppms-acs, 2004:Barbic:RSI, 2010:Yin:JLTP, 2013_Schmidt_RevScInstrum, 2015:Yonezawa:RSI, 2017:Amann:RSI} Areas of interest concern, for instance, automated measurement systems,\cite{1989:Deutz:RSI} set-ups for specific material systems\cite{1995:Nikolo:AJP} or ultra-low temperatures,\cite{2010:Yin:JLTP} susceptometers representing modular add-ons for commercial apparatus,\cite{1995:Nikolo:AJP,2015:Yonezawa:RSI,2017:Amann:RSI} as well as miniaturized suscpetometers,\cite{1997_Pfleiderer_RevSciInstrum} e.g., for use in high pressure cells. 

In this paper we report the design of an AC susceptometer that is optimized for measurements under arbitrary orientation of an applied static field, in particular transverse fields.\cite{2018_Rucker_PhD} The set-up is conceived to be rather compact such that it fits conventional cryogenic systems. The main challenges addressed in the design we report concern excellent thermal coupling, negligible heating effects, and tiny background signal.  The set-up comprises a body made of oxygen-free high-purity copper (OFHC) with sapphire components providing excellent thermal anchoring. This way the amount of metallic material in the immediate vicinity of the primary coil has been minimized, reducing eddy-current heating and parasitic signal contributions. The mechanical support  between the primary and secondaries, implemented by virtue of a sapphire tube, minimizes in particular parasitic signal contributions that originate in vibrations of the primary under transverse magnetic field components. An excellent performance of the susceptometer has been observed in various experimental studies,\cite{2018_Rucker_PhD} illustrated in the following in terms of data recorded at the transverse field Ising quantum phase transition in LiHoF$_4$.
  
\section{Design \& Construction}

\subsection{Longitudinal and Transverse Susceptibility}

The response of the magnetization of a sample in a static magnetic field, $\boldsymbol{H}$, to small changes of the field, $\partial \boldsymbol{H}_\text{AC}\,(\omega)$, may be expressed by the magnetic susceptibility tensor, $\chi_{ij}$. This quantity describes the change of the magnetization component \textit{M}$_i$ in response to a small change of the magnetic field, $\partial H_{\text{AC}, j}\,(\omega)$, along the spatial direction $j$, notably 
\begin{equation} \chi_{ij}(\textbf{\textit{H}}, T, \omega) = \frac{\partial M_i(\textbf{\textit{H}}, T)}{\partial\textit{H}_{\text{AC},j}\,(\omega)} \label{eq:chi_gen}\end{equation}
where, $T$ is the sample temperature, $\omega$ is the frequency, and the indices $i$ and $j$ denote the three spatial directions $x$, $y$ and $z$. 
In general, the direction of the static magnetic field, $\boldsymbol{H}$, and the oscillating excitation field, $\boldsymbol{H}_\text{AC}$, are independent of each other. 

Denoting the direction of the static field with the index~$k$, i.e., $\textit{\textbf{H}} = \left|\textit{\textbf{H}}\right|\cdot\hat{e}_k = H_{k}$, and restricting $\chi_{ij}$ to diagonal elements, $i=j$, two basic configurations are of particular interest. 
First, the so-called longitudinal susceptibility, $\chi_\parallel$, which corresponds to an AC excitation, ${H}_{\text{AC}, j}$, parallel to the static field ${H}_{k}$, i.e., 
\begin{equation}
\label{eq:long}
\chi_\parallel = \chi_{ii}(H_{k}, T, \omega) = \frac{\partial M_i(H_\mathbf{k\parallel i}, T)}{\partial\textit{H}_{\text{AC},i}\,(\omega)}
\end{equation}
This is contrasted by the so-called transverse susceptibility, $\chi_\perp$, for which the AC excitation, ${H}_{\text{AC}, j}$, is perpendicular to the static field ${H}_{k}$, namely 
\begin{equation}
\label{eq:trans}
\chi_\perp = \chi_{ii}(H_{k}, T, \omega) = \frac{\partial M_i(H_\mathbf{k\perp i}, T)}{\partial\textit{H}_{\text{AC},i}\,(\omega)}.
\end{equation} 
As a remark on the side, we note that the off-diagonal elements of the susceptibility tensor ($i\neq j$)  in general do not vanish. These components, however, are expected to be negligibly small in most materials and are not considered in this paper.


\subsection{Design and Dimensions}

The design of the susceptometer reported in this paper is based on the mutual inductance method,  offering great flexibility in terms of the possible coil configurations.\cite{1989:Deutz:RSI, 1995:Nikolo:AJP,1997_Bitko_PhD, 1997_Pfleiderer_RevSciInstrum, ppms-acs, 2004:Barbic:RSI, 2010:Yin:JLTP, 2013_Schmidt_RevScInstrum, 2015:Yonezawa:RSI, 2017:Amann:RSI, 2018_Rucker_PhD} The perhaps most compact implementation combines a primary coil around a single secondary coil. A counter-wound compensation coil may be added on top of the primary to subtract the vacuum contribution.\cite{1997_Pfleiderer_RevSciInstrum} However, when using such a susceptometer for measurements under transverse fields, a large parasitic background signal is generated.\cite{2012_Hirschberger_Diplom} The origin of this parasitic signal may be traced to the mechanical vibrations generated by the primary coil under AC current flow, when the static field is oriented perpendicular to the coil axis. 

In turn, we used a sapphire tube (outer diameter of 8\,mm) as a very rigid and compact support for the primary and secondaries that minimizes the sensitivity to mechanical vibrations while providing excellent thermal coupling and a vanishingly small magnetic susceptibility.\added{\cite{sapphire}} The pair of secondaries, arranged side-by-side, was carefully attached to the inside of the sapphire tube using GE varnish. The secondaries had 1800 windings made of a 38\,$\mu$m diameter enameled copper wire.\cite{coil-supplier} The inside of the secondaries was covered by a thin layer of Stycast Epoxy to prevent mechanical damage when inserting the sample. The primary coil was wound on the outside of the sapphire tube. It had 1485 windings made of a 100$\,\upmu$m enameled copper wire. 

\begin{figure}[htb!]
\centering
\includegraphics[width=\columnwidth]{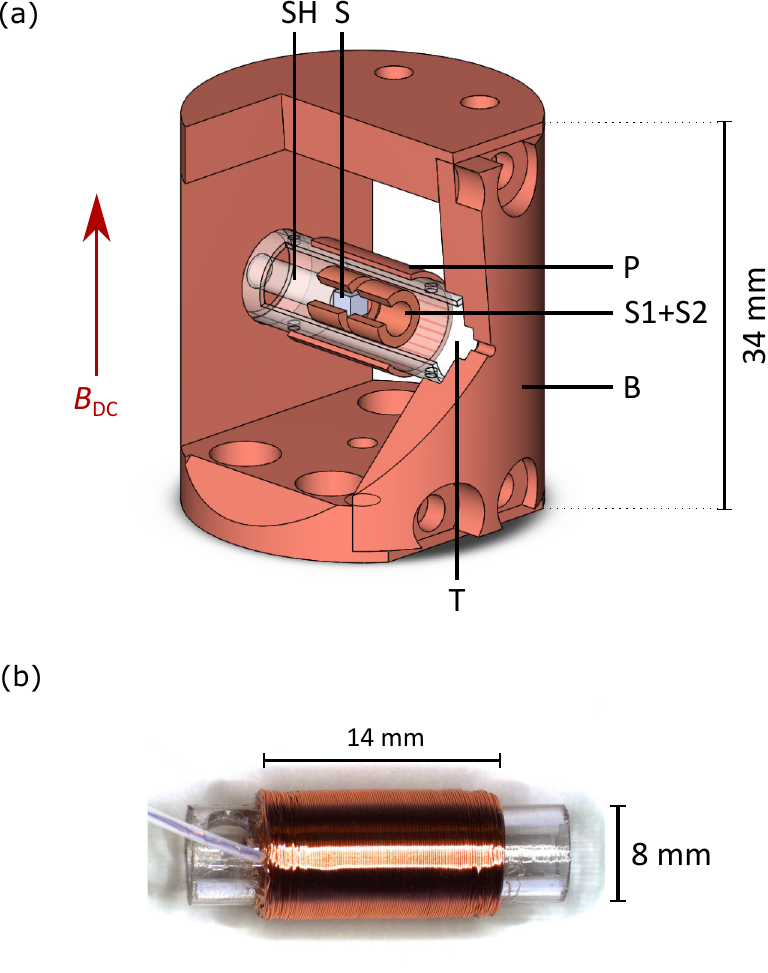}
\caption{Assembled susceptometer described in this paper. (a)~Schematic depiction of the susceptometer. The primary coil (P) is wound onto a sapphire tube (T) inside of which a pair of balanced secondaries (S1+S2) is mounted. The sample (S) is attached to a small sapphire rod (SH) which is inserted into one of the secondaries. The sapphire tube is pressed into an oxygen-free high-purity copper (OFHC) body (B) and additionally fixed by GE varnish. The static field $B_\text{DC}$ may be applied along any direction, say in a vector magnet, notably perpendicular to the susceptometer axis and thus excitation field generated by the primary. (b) Photograph of the sapphire tube with mounted primary coil. }
	\label{fig:suszeptometer}
\end{figure}

The assembly of primary and secondaries on the sapphire tube was mounted into a Cu body, featuring an outer diameter of $d = 30$\,mm and a height of $h =34$\,mm. These dimensions are compatible with conventional cryogenic apparatus and superconducting magnet systems. A schematic depiction of the set-up is shown in Fig.~\ref{fig:suszeptometer}~(a). A small sapphire rod was attached to the copper body using GE varnish and served as a thermal link between the sample and the OFHC Cu body, while minimizing the background signal as well as the effects of eddy currents. One end of the sapphire rod was prepared with a small concentric recession to which the sample was attached using GE varnish. Additionally, the fixed mounting point of the sapphire rod in the Cu body reduced the mechanical cross-coupling between the primary coil and the secondaries. 

A photograph of the sapphire tube with the primary is shown in Fig.\,\ref{fig:suszeptometer}\,(b); the secondaries inside the sapphire tube may not be seen. The arrangement permitted application of static magnetic fields under arbitrary directions, most importantly perpendicular to the axis of the primary with the parasitic background signal being massively reduced. After insertion of the susceptometer into the Cu body, the pair of secondaries was additionally balanced in terms of small changes of the coil position inside the excitation field of the primary coil, before being glued into the final position. This way, a signal to background ratio of $ \approx 10^{4}$ was achieved. Note that the resolution of the susceptometer in this setup is restricted only by the background signal arising from small mis-balancing of the secondaries and small perturbations of the copper body, which exceed common noise sources (such as thermal noise) by several orders of magnitude. The background signal shows no significant temperature or frequency dependence in the temperature and magnetic field range investigated here. 

\begin{figure}[htb!]
\centering
\includegraphics[width=\columnwidth]{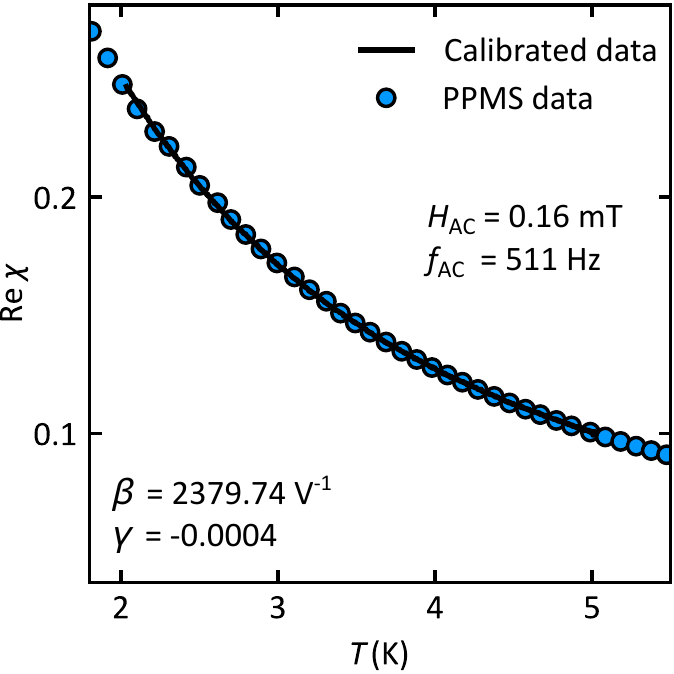}
\caption{AC susceptibility in zero magnetic field of single crystal LiHoF$_4$ as recorded in a PPMS (blue dots) and the compact AC susceptometer (black line) described in this paper. The induced voltage in the susceptometer was scaled by a factor $\beta = 2379.74$V$^{-1}$ and shifted by an offset $\gamma = -0.0004$ to match the PPMS data.}
\label{fig:Sus_Calib}
\end{figure}


\subsection{Operation and Calibration}

For the test measurements reported in the following the assembly of Cu body with susceptometer was firmly attached to the tail of an Oxford Instruments JT dilution refrigerator as operated in a superconducting magnet system. Unfortunately the base temperature of the dilution refrigerator was limited to $\sim69\,{\rm mK}$; we suspect problems with the JT stage or some other unidentified technical source as the cause of the limitation of the base temperature. The base temperature remained unaffected by the susceptomter.

The temperature of the susceptometer was monitored by means of three calibrated RuOx thermometers. The first sensor was used as control thermometer and screwed to the mixing chamber, the second sensor was attached to the copper body close to the sample holder using a flat silver band and GE varnish. The third thermometer was screwed to the bottom of the susceptometer. Despite the tiny cooling power of the dilution unit ($\sim25\,{\rm \mu W}$ at 0.1\,K) the entire set-up as monitored by the thermometers came into thermal equilibrium without any delay for the accessible cooling rates smaller than $\Delta T < 50\,\frac{\text{mK}}{\text{min}}$ at all temperatures. This provided direct evidence for the excellent thermal anchoring of all components. 

A Keithley 6221 AC current source was used to generate typical AC excitation currents of the order $I = 100\,\upmu$A. Considering the coil geometry, this resulted in an amplitude of the excitation field  $B_\text{AC} \approx 0.16$\,mT. The susceptometer was tested for excitation frequencies between $f_\text{AC} = 10 $\,Hz and $f_\text{AC} = 5$\,kHz. A Signal Recovery SR830 digital lock-in amplifier was used to detect the difference of the induced voltages in the secondaries. The  signal may be calculated by the relation
\begin{equation}
U_{\text{ind}} = U_{\text{ind}}^{\text{vac}} + U_{\text{ind}}^{\text{sample}} 
= \mu_0 H_0 \omega A N_s \sin(\omega t)(1 + \chi \mathit{f}) 
\end{equation}
where $B_\text{AC}= \mu_0H_0$ is the excitation amplitude, $\omega$ is the excitation frequency, $A$ is the cross-section of the secondaries, $N_s$ is the number of windings of the secondaries, $\chi$ is the susceptibilty of the sample, and $f$ the volume fraction of the sample inside one of the secondaries. 

The signal of the susceptometer was calibrated by means of a reference material. For this purpose the signal of a LiHoF$_4$ single crystal was at first determined quantitatively in a Quantum Design physical properties measurement system (PPMS). The susceptibility was recorded for temperatures between 2\,K and 5\,K in the PPMS, as well as with the susceptometer described in this paper. The induced voltage signal $U_\text{ind}$ of the susceptometer was scaled by a factor $\beta$ to fit the PPMS data in SI units according to
\begin{equation}
\chi_\text{PPMS} = U_\text{ind}\cdot \beta + \gamma
\end{equation} 
where the additional parameter $\gamma$ accounted for a tiny offset due to imperfect balancing of the secondaries. Shown in Fig. \ref{fig:Sus_Calib} is the calibrated susceptibility of the LiHoF$_4$ sample as measured in the PPMS and the susceptometer described in this paper. Both measurements were recorded for an excitation amplitude $B_\text{AC} \approx 160\,\upmu$T and a frequency of $f_\text{AC} = 511$\,Hz. The scaling factor was $\beta = 2379.74$\,V$^{-1}$ and the offset $\gamma = -4\cdot10^{-4}$. The small value of $\gamma$ showed that the secondaries were balanced well. 


\begin{figure}[h!]
\centering
\includegraphics[width=\columnwidth]{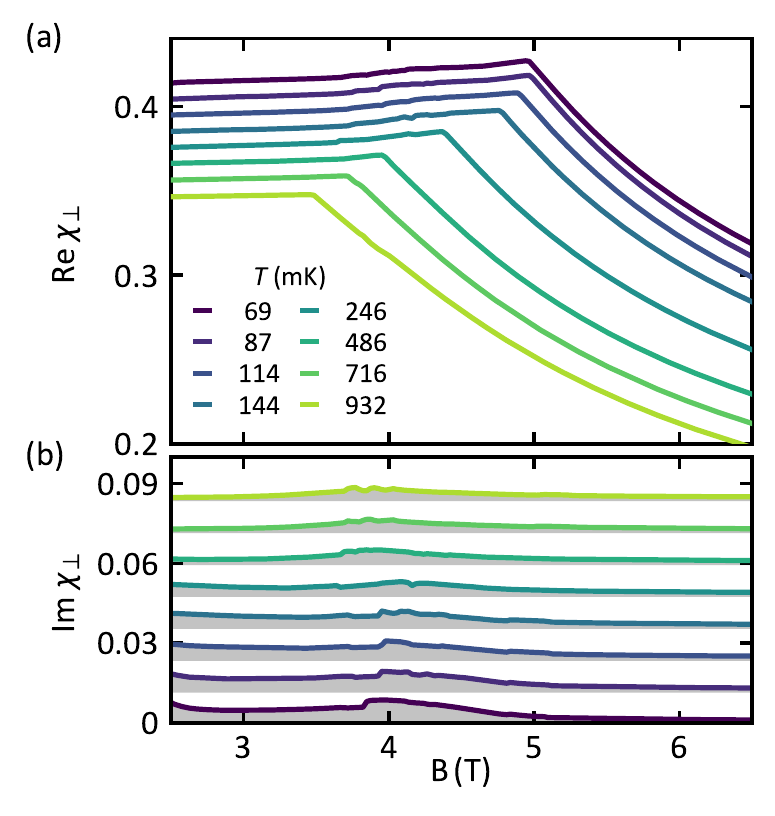}
\caption{Transverse susceptibility of a spherical single-crystalline sample of LiHoF$_4$ as a function of magnetic field. (a) Real part of the susceptibility, \rchi, as a function of magnetic field $B$ at different temperatures $T$. The ferromagnetic to paramagnetic phase transition results in a kink with a maximum critical field $B_{c} \approx 5\,$T for the lowest temperature investigated. (b) Imaginary part of the AC susceptibility, \ichi{}, as a function of magnetic field. Tiny contributions in the ferromagnetic state provide evidence of small dissipative processes. In both data sets, a constant offset has been added for better visibility.}
\label{fig:Data}
\end{figure}

\section{Performance}

The low-temperature properties of the Ising ferromagnet LiHoF$_4$ at the transverse field quantum phase transition were measured in order to illustrate the excellent performance of the susceptometer. In the literature, the properties of LiHoF$_4$ are portrayed as a text book example of a quantum phase transition.\cite{1996_Bitko_PhysRevLett, 2005:Ronnow:Science, 2007_Legl_PhD, 2011_Sachdev_Book} The investigation of such model systems offers important insights into fundamental notions of condensed matter systems such as the nature of the elementary excitations near quantum criticality and allows comparison with well-understood theoretical predictions. In LiHoF$_4$, the QPT is induced by a strong magnetic field, applied perpendicular to the Ising-axis. In turn, the easy-axis susceptibility under transverse field, which corresponds to the transverse susceptibility as defined in Eq. \ref{eq:trans}, represents the key quantity that characterizes the transition. 

Shown in Fig. \ref{fig:Data} are typical experimental data as recorded with the susceptometer. Fig.\,\ref{fig:Data}\,(a) displays the real part, \rchi{}, of the transverse susceptibility and illustrates the performance of the susceptometer as a function of transverse magnetic field at different temperatures down to $T=69$\,mK. The ferromagnetic to paramagnetic phase transition is accompanied by a kink for all temperatures investigated, approaching an extrapolated critical field, $B_c \approx 5\,T$, for the lowest temperature studied. The resulting magnetic phase diagram is in perfect agreement with previous studies.\cite{1996_Bitko_PhysRevLett} 

The characteristic $B^{-1}$ dependence of the susceptibility as a function of transverse field for $B\geq B_{c}$, as well as the demagnetization plateau at $B\leq B_{c}$ indicate excellent quantitative operation of the susceptometer. No additional background contribution was observed for the field range investigated. Fig.\,\ref{fig:Data}\,(b) displays the imaginary part of the susceptibility, \ichi{}, as a function of transverse magnetic field. In agreement with previous studies, no significant contribution to the imaginary part of the susceptibility was observed over the entire range of magnetic fields and temperatures. A tiny contribution to \ichi{} by the sample may be noticed around $B\approx 4$\,T. A technical origin of this feature due to the susceptometer may be ruled out, as no such signal was observed in other studies using the same susceptometer as well as different materials. More subtle processes associated with the quantum phase transition are likely to be the cause of these contributions. They are, however, beyond the scope of the work reported here.
 
\begin{figure}[t]
\centering
\includegraphics[width=\columnwidth]{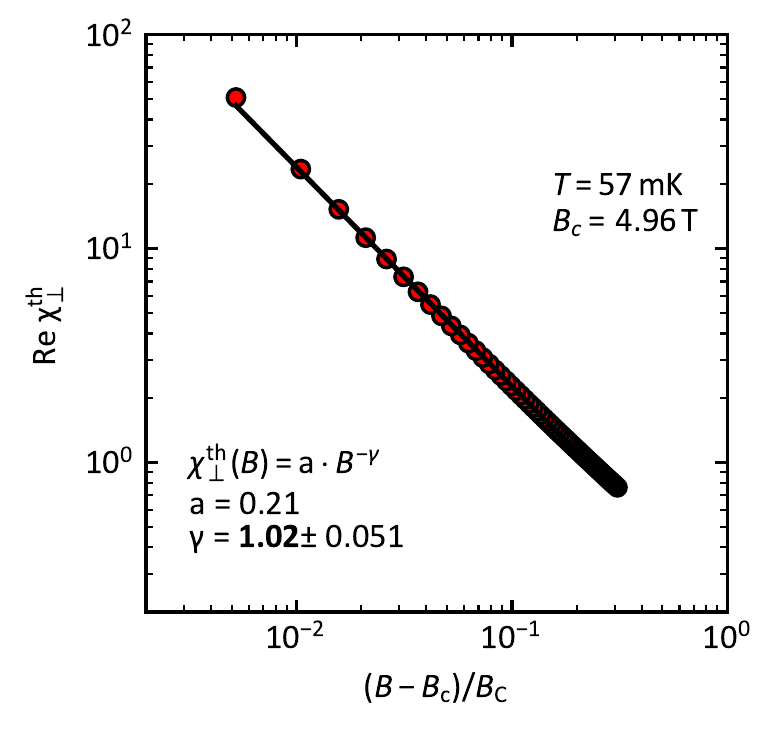}
\caption{Real part of the AC susceptibility corrected for demagnetization effects, Re\,$\chi_\perp^\text{th}$, as a function of reduced magnetic field $b = (B-B_c)/B_c$. A mean-field critical exponent of $\gamma = 1.02\pm0.051$ is observed in close proximity to the phase transition, in agreement with theoretical predictions and previous studies.\cite{1996_Bitko_PhysRevLett}}
	\label{fig:Data_2}
\end{figure}

Shown in Fig.\,\ref{fig:Data_2} is the critical exponent at the QPT of LiHoF$_4$ inferred from the experimental data. Here, the theoretically expected real part of the susceptibility, Re\,$\chi^\text{th}_\perp$, is shown as a function of the reduced magnetic field ${(B-B_{c})}/{B_c}$. This susceptibility value is given by the experimentally determined value \rchi, as corrected for demagnetization effects using \mbox{$\chi_\perp^\text{th} = [{(\chi_\perp)^{-1}-(\chi_\perp^\text{max})^{-1}}]^{-1}$}, where $\chi_\perp^\text{max}$ represents the plateau of the susceptibility for $B\leq B_{c}$.\cite{1978_Beauvillain_PhysRevB} Here we found a critical exponent $\gamma = 1.02\pm 0.051$ in excellent agreement with the theoretically predicted mean-field exponent $\gamma =1$, as well as the experimental study by Bitko \textit{et al.} \cite{1996_Bitko_PhysRevLett,1997_Bitko_PhD}, who reported a critical exponent $\gamma = 1.07\pm 0.11$. The small difference of the value we observe is empirically consistent with the uncertainties due to the tiny background contributions of the empty susceptometer.

\section{Conclusions}

In conclusion, we reported a compact design of a susceptometer suitable for measurements under static magnetic fields applied along arbitrary directions, in particular large transverse magnetic fields. The signal of the susceptometer was calibrated by means of a comparison with data recorded in a conventional PPMS in zero magnetic field. To characterize the performance in the millikelvin regime, the transverse field tuned quantum phase transition in LiHoF$_4$ was revisited, where data were found to be in excellent agreement with the literature.\cite{1996_Bitko_PhysRevLett} This establishes that the susceptometer displays a very large ratio of signal to background above $\sim10^{4}$ at temperatures as low as $\sim 57$\,mK under transverse magnetic fields up to $B = 6.5$\,T, representing the lowest temperatures and largest fields accessible with our cryogenic equipment.

The overall dimensions of the set-up described here are suitable for typical cryogenic magnet-systems, where the excellent thermal anchoring of the sample permits measurements well into the milli-Kelvin regime as demonstrated for temperatures as low as $T\approx 57\,$mK. Taken together the susceptometer described here is in particular suitable as a module for use in a Quantum Design ppms system,\cite{ppms-acs} and dilution refrigerators with comparatively low cooling powers such as the system used for the experimental tests reported in this paper.

\section{Acknowledgements}
Discussions with and support of A. Bauer A. Chac\'on, C. Duvinage, M. Hirschberger and S. Legl are gratefully acknowledged. Financial support through DFG TRR80 (project E1, F2 and F7), SPP 2137 (Skyrmionics), and ERC advanced grant TOPFIT (291079) is acknowledged. This project has received funding from the European Research Council (ERC) under the European Union's Horizon 2020 research and innovation programme (grant agreement No 788031, ExQuiSid). FR acknowledges financial support of the TUM Graduate School.


%

\end{document}